\documentclass[10pt]{amsart}
\usepackage{amsmath,amssymb}
\usepackage[dvips]{epsfig}
\newtheorem{theorem}{Theorem}[section]
\newtheorem{lemma}[theorem]{Lemma}
\newtheorem{corollary}[theorem]{Corollary}
\newtheorem{proposition}[theorem]{Proposition}
\theoremstyle{definition}

\theoremstyle{remark}
\newtheorem{example}[theorem]{Example}

\allowdisplaybreaks
\numberwithin{equation}{section}
\newcommand{\rnc}{\renewcommand}
\newcommand{\nc}{\newcommand}
\newcommand{\be}{\boldsymbol{\epsilon}}
\newcommand{\bd}{\boldsymbol{\delta}}
\nc{\pq}{\left[\boldsymbol{e}\right]}
\nc{\mpq}{\left[^{-\boldsymbol{p}}_{-\boldsymbol{q}}\right]}
\nc{\pqt}{\left[^{\tilde{\boldsymbol{p}}}_{\tilde{\boldsymbol{q}}}\right]}
\nc{\de}{\left[^{\delta}_{\epsilon}\right]}
\newcommand{\res[1]}{\operatornamewithlimits{Res}_{#1}}

\nc{\no}{\nonumber} \nc{\IM}{\hbox{\mathrm{Im}}\,}
\nc{\A}{\mathcal{A}}
\nc{\ID}{{\mathbf 1}} \nc{\RE}{\mathrm{Re}\,}
\nc{\Id}{\mathrm{Id}\,}
\nc{\C}{{\mathbb C}}
\nc{\T}{{T}}
\nc{\Cp}{{\mathbbC\rm I\!P}}
\nc{\Z}{\mathbb{Z}}
\nc{\bv}{\boldsymbol{v}}
\newcommand{\Rn}{{\rm I\!R}}
\nc{\e}{\mbox{e}}
\rnc{\i}{\imath}
\nc{\Uc}{\boldsymbol{\mathcal{U}}}
\nc{\Ac}{\mathcal{A}}
\nc{\Bc}{\mathcal{B}}
\nc{\Cc}{\mathcal{C}}
\nc{\tAc}{\widetilde{\mathcal{A}}}
\nc{\tBc}{\widetilde{\mathcal{B}}}
\nc{\tCc}{\widetilde{\mathcal{C}}}
\nc{\p}{\bf p}
\nc{\q}{\bf q}
\rnc{\d}{\mathrm{d}}
\nc{\lb}{\lambda}
\rnc{\a}{\alpha}
\rnc{\b}{\beta}
\nc{\hS}{\hat{S}}
\nc{\g}{\it{g}}
\nc{\Si}{\Sigma}

\newcommand{\ra}{\rightarrow}

\begin{document}
\title[Thomae  formulae for
singular $Z_N$ curves]
{Thomae type formulae for singular $Z_N$ curves}
\author[Enolski]{V.Z. Enolski}
\address{Institute of Magnetism NASU, Vernadsky blvd. 36, Kyiv-142, 
Ukraine}\email{vze@ma.hw.ac.uk}

\author[Grava]{T.Grava}
\address{SISSA, via Beirut 2-4 340104 Trieste, Italy\\
E-mail: grava@sissa.it}
\begin{abstract}
We  give an elementary and rigorous proof of the Thomae type formula
for the singular  curves $\mu^N=\prod_{j=1}^{m}(\lb-\lambda_{2j})^{N-1}
\prod_{j=0}^{m}(\lb-\lambda_{2j+1})$.
To derive the Thomae formula we use the traditional variational method
which goes back to Riemann, Thomae and Fuchs. An important step of the proof is the use
 of the Szeg\"o kernel computed explicitly in algebraic form
for non-singular $1/N$-periods. The proof inherits principal points of
Nakayashiki's proof \cite{na97}, obtained for non-singular $Z_N$ curves.
\end{abstract}

\thanks{VZE wish to thank Mittag-Leffler Institute for support 
within program ``Wave motion'' in December 2005 when the final version 
of this paper was made. TG acknowledges support of the European Science Foundation Program
MISGAM.}
\keywords{Thomae type formula, Riemann surfaces, kernel forms,  
Rauch variational formulas, nonsingular characteristics. MSC:
 35Q15, 30F60, 32G81}
\maketitle

\section{Introduction}
The original Thomae formula \cite{th69} expresses the zero values
of the Riemann $\theta$-functions with half integer characteristics
as functions of the branch points of the hyperelliptic curve
$\mu^2=\prod_{i=1}^{2m}(\lb-\lambda_i)$, that is
\begin{eqnarray}
&& \theta [\boldsymbol{e}](\boldsymbol{0};\Pi)^8
=\Big(\frac{\det \mathcal{A} }{ (2\pi
i)^{m-1}}\Big)^4
\prod_{k<l}(\lb_{i_k}-\lambda_{i_l})^2(\lb_{j_k}-\lambda_{j_l})^2,
\label{thomaeoriginal}
\end{eqnarray}
where $\boldsymbol{e}$ is a non-singular even half-period
corresponding to the partition of the branch points
$\{1,\cdots,2m\}=\{i_1<\cdots<i_m\}\cup\{j_1<\cdots<j_m\}$.
The  Riemann period matrix $\Pi$ and the matrix of $\alpha$-periods
$\mathcal{A}=(\oint_{\alpha_i}\lb^{j-1}/\mu)_{1\leq i,j\leq m-1}$
are computed in a canonical homology basis  $(\boldsymbol{\alpha},
\boldsymbol{\beta})$.

The modern interest in the Thomae formulae was initially
stimulated by the finite-gap integration  of KdV  and
KP type equations where, according to the Dubrovin-Novikov program
of "effectivization of the finite-gap solutions",  it was necessary
to express winding vectors in the Its-Matveev formula~\cite{im75}
in terms of $\theta$-constants (see e.g.~\cite{du81,mu83}). In
this context Thomae formulae were recently considered in
\cite{er05}.

Thomae formulae were used to describe the moduli space of
hyperelliptic curves with level two structure in terms of theta
constants \cite{mu83}. They give a generalization of the $\lambda$
function of an elliptic curve
 \cite{mu83} (Umemura's appendix) and \cite{farkas96}. In~\cite{me91}
 they were combined with the arithmetic-geometric
mean to obtain criteria for the reducibility of ultraelliptic Jacobians.
Thomae formulae also appear in a wider context of modern
research on integrable systems, in particular in
the Riemann-Hilbert problem and the 
associated Schlesinger equations (see, e.\,g.,\cite{kk98}).
In this set of problems the Thomae formulae were used
to give explicit expression for the $\tau$-function of the 
Schlesinger equation associated  with hyperelliptic curve
\cite{kk98}, general curve with simple branch points
\cite{kor01} and non simple  branch points \cite{eg04}.
Thomae formulas are also used  to compute
action variables in conditions of complete integrability 
(e.\,g.~\cite{drvw01}).
In the conformal field theory Thomae formulae became relevant  after
the work of Knizhnik \cite{knizh89} who expressed correlation 
functions of the multi-loop
string amplitude in terms of $\theta$-constants of a Riemann surface with $Z_N$ symmetry.
A solution  of Knizhnik-Zamolodchikov equations in terms
of $\theta$-constants was also obtained in \cite{sm93} on the basis of
Thomae formulae.

After the classical Thomae paper \cite{th69} the derivation of
Thomae formulae in hyperelliptic case was given in many places:
Fuchs \cite{fuc71}, Bolza \cite{bo99}, Fay \cite{fa73},
Mumford \cite{mu83}. But only recently Bershadsky and Radul
\cite{br87,br88}  discovered a generalization of Thomae formula for
${ Z}_N$ curve $\mu^N=\prod_{i=1}^{Nm}(\lambda-\lambda_i)$ and gave
a heuristic proof of them on the basis of path integral formulation
of  the conformal field theory. Afterwards Nakayashiki \cite{na97} developed a
 rigourous  derivation of these formulae in the frame of
the classical methods.

In this paper we derive the Thomae formula for the singular  curve
\begin{equation}
 \mu^N=\prod_{j=1}^{m}(\lambda-\lambda_{2j})^{N-1}\prod_{i=0}^m
(\lambda-\lambda_{2i+1}),\quad N>1 \quad N\in\mathbb{N},
\label{zvcurve}\end{equation}
which has $Z_N$ symmetry and it will be called   singular $Z_N$ curves.

The modular properties of certain family of  curves  (\ref{zvcurve}) 
 were investigated by Burhardt \cite{bur93},
Hutchinson \cite{hut02}, more recently by Shiga \cite{shiga88}
and Koike \cite{koike03}, Diez \cite{diez91} and others.
This curve (\ref{zvcurve})  appeared in Zverovich \cite{zv71} as
the curve related to a solvable Riemann-Hilbert problem with
quasi-permutation monodromy matrices. Explicit solution of this problem was
derived in our recent paper \cite{eg04} which stimulated  the present investigation.

Our proof of Thomae formulae 
 goes up to the original Thomae paper \cite{th69} and inherits
principal steps of the Nakayashiki's proof 
\cite{na97} for non singular $Z_N$ curves.
The non singular $Z_N$ curves are invariant under the group generated
by the permutations of branch points, while the singular  $Z_N$ curves
are not. This is the main difference with the case treated by Nakayashiki.

The main steps of our proof are the  explicit algebraic formula for the  
Szeg\"o kernel associated with non-singular $1/N$-period and the 
explicit formula for the canonical bimeromorphic differential kernel.
 The above kernel functions can be realized in $\theta$-functional form,
 and in algebraic forms. 
In the former case the theory was developed by Fay  \cite{fa73}, in the latter case the  theory has  originated in the work of Klein \cite{kl86,kl88} and later it  was  developed by Baker \cite{ba97,ba07} and also 
\cite{hs66},\cite{bel97b} and for nonsingular  $Z_N$ curves by  Nakayashiki 
\cite{na97}. 
The comparison of these representations leads to a 
certain relation which is the important point of the proof.  
We also use the notion of  variation of branch points by means of 
Rauch formulae \cite{ra59},\cite{kor01}.

The paper is organized as follows. In  Section 2 we are fixing the 
notations and remind basic notions from the theory of algebraic curves
which we need for the foregoing development. In the  Section 3 we
describe characteristics of Abelian images of branch points. 
We introduce in this section families of non-special
divisors supported on  branch points and describe corresponding 
$1/N$-periods. We describe canonical bimeromorphic differential and 
Szeg\"o kernels in the Section
4 and develop their expansions. In the Section 5 we briefly discuss
Kleinian formulation of the kernel forms. All these results  are
summarized in the Section 6 where the formulation and proof of the
Thomae formula is given. We briefly discuss further perspectives in
the last Section.

\section{The curve, its homologies,
differentials, $\theta$-functions and variations}
Consider the Riemann surface
$\mathcal{C}$ of the curve
\begin{align}
\mu^N&=p(\lb)q(\lb)^{N-1},\label{CNM}\\
 p(\lb)&=\prod_{k=0}^{m}(\lb-\lambda_{2k+1}),\quad
q(\lb)=\prod_{k=1}^{m}(\lb-\lambda_{2k}).\label{pq}
\end{align}
 The curve (\ref{CNM})   has branch points $P_k=(\lb_k,0)$, $k=1,\dots,2m+1$ and $P_{2m+2}=P_{\infty}=(\infty,\infty)$ and  singularities
at the points $P_{2k}$, $k=1,\dots,m+1$. These singularities 
can be easily resolved \cite{mi95} to give rise to a compact Riemann
surface which we still denote by $\mathcal{C}_{N,m}$.
The genus $g$ of $\mathcal{C}_{N,m}$  is equal to $(N-1)m$.

The projection $(\lb,y)\ra \lb$, (which we still denote by $\lb$)
defines $\mathcal{C}_{N,m}$ as a $N-$sheeted
covering of the complex plane   $\mathbb{CP}^1$ branched over the points
$P_k$, $k=1,\dots 2m+2$. The
pre-image of a non-branch  point $\lb\in \mathbb{CP}^1$
consists of $N$ points. The $N$-cyclic automorphism $J$ of
$\mathcal{C}_{N,m}$ is given by the action 
$J:(\lb,y)\ra (\lb,\rho y)$, where $\rho$ is the $N$-primitive root  of unity, 
namely $\rho=\e^{\frac{2\pi \i}{N} }$.
For $P$ in a neighbourhood $U_R$ of  the point
$R\in\mathcal{C}_{N,m}$, a local coordinate $x(P)$, with $x(R)=0$, 
 is  the function defined by
\begin{equation}
\label{localN}
x(P)=\begin{cases} \lambda(P)-\lambda(R), &\text{if}\quad R\quad
\text{is an ordinary point, }\\
            \sqrt[N]{ \lambda(P)-\lambda(R)}, &\text{if}
\quad R=P_k,\;k=1,\dots,2m+1,\\
            \frac{1}{\sqrt[N]{\lambda(P)}},&\text{if}\quad R=P_{\infty}.
\end{cases}
\end{equation}

\begin{figure}[htb]
\centering
\mbox{\epsfig{figure=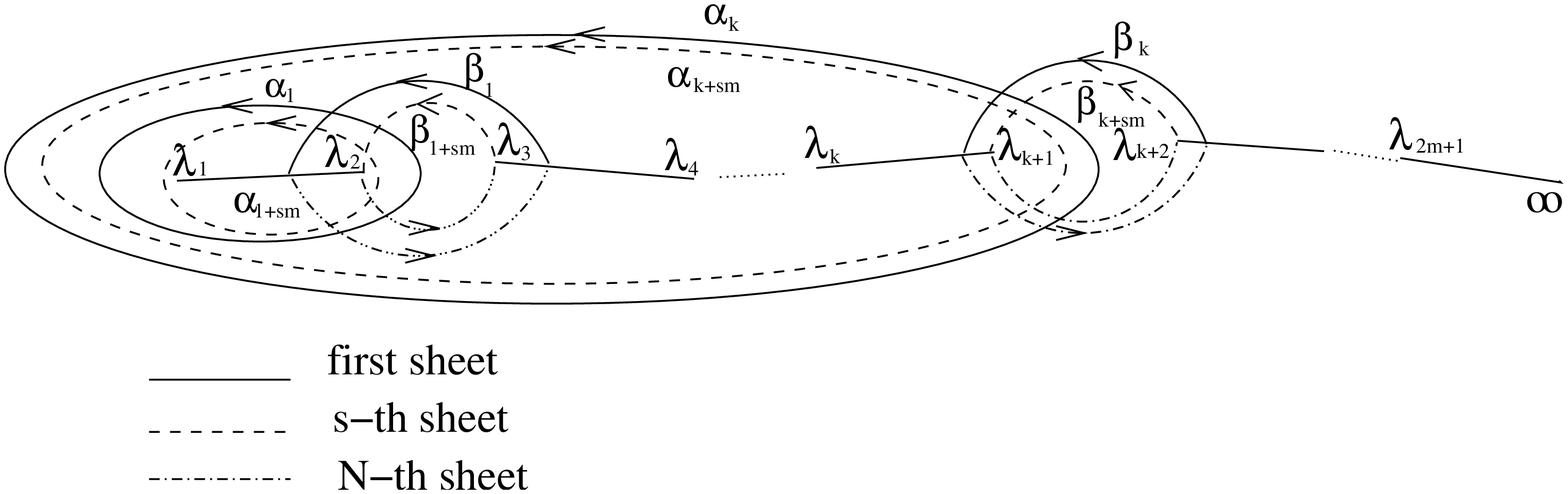,width=0.95\textwidth}}
     \caption{The homology basis. \label{fig2}}
\end{figure}

The canonical homology basis,
$(\alpha_1,\ldots, \alpha_{(N-1)m}; \beta_1,\ldots,
\beta_{(N-1)m})\in H(\mathcal{C},\mathbb{Z})$
of the curve $\mathcal{C}$ is
shown in the Figure~\ref{fig2}.
Namely the cycles  $\alpha_{k+sm}$,  $k=1,\dots,m$ lie on the
$(s+1)-$th sheet, $s=0,\dots,N-2$ and encircle anti-clockwise the cuts
$\cup_{j=1}^k(\lb_{2j-1},\lb_{2j})$. The cycles $\beta_{k+sm}$,
$k=1,\dots, m$, $s=0,\dots,N-2$,
emerges on the $(s+1)-$th sheet through the point
$(\lambda_{2k+1},0)$  pass anti-clockwise to the $N$-th
sheet through the point $(\lambda_{2k},0)$  and return to the initial point
through the $N$-th sheet. We remark that on  Figure 1, when $N>3$, the  $\beta$-cycles placed
from the  second  to the $(N-2)$th sheet  should intersect the cuts only
in the  branch  points. It is possible to drop this requirement and plot
appropriate number of loops around each branch point.

The action of the automorphism  $J$ on the basis of cycles is given by
\begin{align*}
&J(\alpha_{i+sm})=\alpha_{i+(s+1)m},~~~~i=1,\dots,m,~~~s=0,\dots,N-3,\\
&J(\alpha_{i+(N-2)m})=-\sum_{s=0}^{N-2}\alpha_{i+sm},\quad i=1,\ldots,m,\\
&J(\beta_{i+sm})=\beta_{i+(s+1)m}-\beta_i, \;\;s=0,\dots,N-3,
\\& J(\beta_{i+(N-2)m})=-\beta_i,\quad i=1,\ldots,m.\notag
\end{align*}
The  basis of canonical holomorphic differentials reads
\begin{equation}
\label{holo}
\mathrm{d}u_{j+sm}(P)=\dfrac{\lb^{j-1}q(\lb)^{s}}{\mu^{s+1}}\mathrm{d}\lb,
\quad j=1,\dots,m,\;\;s=0,\dots,N-2.
\end{equation}
The induced action of $J$ on the holomorphic differential is given by
\[
J(\mathrm{d}u_{j+sm}(P)):=\mathrm{d}u_{j+sm}(J(P))=
\dfrac{\lb^{j-1}q(\lb)^{s}}{\rho^{s+1}\mu^{s+1}}\mathrm{d}\lb,
\quad j=1,\dots,m,\;\;s=0,\dots,N-2,
\]
where $\rho$ is the $N$-th root of unity.
The $(N-1)m\times(N-1)m $ matrices  $\A$ of  $\alpha$-periods and
$\mathcal B$ of  $\beta$-periods are expressible in terms of
$m\times m$-matrices
\begin{equation}
\label{As}
\A_{s+1}=\left(\oint_{\alpha_{j}}\d u_{k+ms}\right)_{k,j=1,\ldots,m},
\;\quad
\mathcal{B}_{s+1}=\left(\oint_{\beta_j}\d u_{k+ms}\right)_{k,j=1,\ldots,m},
 \;\quad s=0,\dots,N-2,
\end{equation}
in the following way. Let us introduce the $(N-1)m\times (N-1)m$ dimensional
 matrices
\begin{align}
\label{RA}
&\mathcal{R}_{\mathcal{A}}=\left(\dfrac{\rho^{-i(k-1)}-\rho^{-ik}}{1-\rho^{-i}}\right)_{i,k=1,\dots,N-1}\otimes 1_{m},\\
\label{RB}
&\mathcal{R}_{\mathcal{B}}=\left(\dfrac{\rho^{-i(k-1)}-\rho^{-i(N-1)}}{1-\rho^{-(N-1)i}}\right)_{i,k=1,\dots,N-1}\otimes 1_{m}.
\end{align}
Then
\begin{align}
\label{Aperiod}
\A&=\left(\oint\limits_{\alpha_j}
\mathrm{d}u_k\right)_{k,j=1,\ldots,(N-1)m}=
\mbox{Diag}(\mathcal{A}_1,\mathcal{A}_2,\dots,
\mathcal{A}_{N-1})\mathcal{R}_{\mathcal{A}},\\
\mathcal{B}&=\left(\oint\limits_{\beta_j}
\mathrm{d}u_k\right)_{k,j=1,\ldots,(N-1)m}=\mbox{Diag}(\mathcal{B}_1,
\mathcal{B}_2,\dots, \mathcal{B}_{N-1})\mathcal{R}_{\mathcal{B}},
\label{Bperiod}
\end{align}
where
$
\mbox{Diag}(\mathcal{A}_1,\mathcal{A}_2,\dots, \mathcal{A}_{N-1}),
\quad\mbox{Diag}(\mathcal{B}_1,\mathcal{B}_2,\dots, \mathcal{B}_{N-1})
$
are  the block diagonal matrices
having as entries the matrices $\mathcal{A}_s$ and $\mathcal{B}_s$,
$s=1,\dots,N-1$, respectively.

The normalized holomorphic differentials $\d\boldsymbol{v}=(\d
v_1,\dots,\d v_{(N-1)m})$ are given as $ \d \boldsymbol{v}= \d
\boldsymbol{u}\mathcal{A}^{-1}$ and the Riemann period matrix $\Pi$,
$$\Pi=\left(\oint_{\beta_{k}}\d v_j\right)_{ j,k=1,\dots,(N-1)m}$$ is
given by
$
\label{periodmatrix}
\Pi=\mathcal{R}_{\mathcal{A}}^{-1}\mbox{Diag}(\mathcal{A}^{-1}_1\mathcal{B}_1,
\mathcal{A}^{-1}_2\mathcal{B}_2,\dots, \mathcal{A}^{-1}_{N-1}
\mathcal{B}_{N-1})\mathcal{R}_{\mathcal{B}}
$
with $\mathcal{R}_{\mathcal{A}}$ and $\mathcal{R}_{\mathcal{B}}$
defined in (\ref{RA})
and (\ref{RB}). The matrix $\Pi$ is necessarily symmetric and has positively defined imaginary part. The space of such matrices is called Siegel half-space $\mathcal{H}_g$. Denote also Jacobi variety of the curve
$\mathrm{Jac}(\mathcal{C})=\mathbb{C}^g/ 1_g\oplus\Pi $.

To complete this Section we recall the important
{\it Rauch variation formulae} \cite{ra59}
for the period matrix $\Pi$ 
\begin{equation}
\label{Rauch1} \dfrac{\partial}{\partial \lb_k}\Pi_{ij}=2\pi \i
\res[\lb=\lb_k]\left\{ \dfrac{1}{(\d \lb(P))^2}\sum_{s=1}^N\d
v_i(P^{(s)}) \d v_j(P^{(s)})\right\},\;\;
\end{equation}
where $i,j=1,\dots,2m,\;\; k=1,\dots,2m+1.$ We remark that in the
case of hyperelliptic curves the formula (\ref{Rauch1})
appeared already in the
Thomae article \cite{th69}. For general surfaces the infinitesimal
variation of Abelian differentials and their periods with respect to Beltrami
differentials is due to Fay \cite{fa92}. 
Korotkin \cite{kor01}  reduced the variation formula to useful form (\ref{Rauch1}).  
The proof of the  variational formula for the holomorphic differentials can be found in \cite{kokkor03}.

\subsection{Theta functions}
Any point $\boldsymbol{e}\in\mathrm{Jac}(\mathcal{C})$
can be written uniquely as
$\boldsymbol{e}=(\be,\bd)\left(^{1_g}_\Pi\right)$,
where $\be,\bd\in\Rn^g$ are the characteristics
of $\boldsymbol{e}$.
We shall use the  notation $[\boldsymbol{e}]=[^{\bd}_{\be}]$.
If $\be$ and $\bd$ are half integer,
then we say that the corresponding characteristics
$[\boldsymbol{e}]$ are half-integer.
The half-integer  characteristics are odd or even, whenever
 $4\langle\bd,\be\rangle$ is equal
to $1$ or $0$ modulo $2$.
Here and below the brackets $\langle\;,\;\rangle$  denote the standard
Euclidean scalar product.
The Riemann $\theta$-function with characteristics $\pq$ is defined in
 ${\mathcal H}_g\times \mathrm{Jac}({\mathcal C})$ as the Fourier series
\begin{align}
\theta\pq(\boldsymbol{z};\Pi)&=\sum_{\boldsymbol{n}\in \mathbb{Z}^g}
\exp(\pi \i\,\langle 
\boldsymbol{n}\Pi+\bd \Pi,\boldsymbol{n}+\bd\rangle+2\pi
\i\,\langle
\boldsymbol{z}+ \be, \boldsymbol{n}+\bd\rangle)\\
&=\theta(\boldsymbol{z}+\boldsymbol{\epsilon}+\boldsymbol{\delta}\Pi;\Pi)
\mathrm{exp}\{2\i\pi\langle \boldsymbol{\delta},(\boldsymbol{z}
+\frac12\boldsymbol{\delta}\Pi )\rangle
+2\i\pi\langle\boldsymbol{\epsilon},\boldsymbol{\delta}
\rangle  \}.\notag
\end{align}
For $\pq=[\boldsymbol{0}]$ we write $\theta[\boldsymbol{0}](\boldsymbol{z};\Pi)=
\theta(\boldsymbol{z};\Pi)$.
The $\theta$-function is an entire in
$\boldsymbol{z}\in\C^g$ and has  periodicity properties:
\begin{align}
\label{periodicity1}
\theta\pq(\boldsymbol{z}+\boldsymbol{m}'+\boldsymbol{m}\Pi;\Pi  )=
\mathrm{exp}\{ -2\i\pi
\langle\boldsymbol{m}',(\boldsymbol{z}+\frac12\boldsymbol{m}'\Pi)
\rangle
-2\i\pi \langle \boldsymbol{\epsilon},
\boldsymbol{m}'\rangle  \} \theta\pq(\boldsymbol{z};\Pi),
\end{align}
where $\boldsymbol{m}$ and $\boldsymbol{m}'$ are integer vectors.

The $\theta$-function with arbitrary
characteristics satisfies  the heat equation
\begin{equation}
\frac{\partial^2}{\partial z_k\partial
z_l}\theta\pq(\boldsymbol{z};\Pi)=
2\imath\pi(1+\delta_{k,l})\frac{\partial}{\partial
\Pi_{kl}}\theta\pq(\boldsymbol{z};\Pi), \quad k,l=1,\ldots,g.
\label{heat}\end{equation}

The zeros of the $\theta$-function are described by the fundamental
{\it Riemann singularity theorem}.
\begin{theorem}
\label{vanishing}
Let $\boldsymbol{e}\in \mathrm{Jac}({\mathcal C})$ be an arbitrary
vector and $Q_0\in\mathcal{C}$- arbitrary point.  
Then the multi-valued function
\[
P\rightarrow \theta\left(\int_{Q_0}^P\d\boldsymbol{v}-\boldsymbol{e};\Pi\right)
\]
has on $\mathcal{C}$ exactly $g$ zeros $Q_1,Q_2,\dots,Q_g$ provided
it does not vanish identically and   
\begin{equation}
\label{vectore}
\boldsymbol{e}=\sum_{i=1}^g\int_{Q_0}^{Q_i}
\d\boldsymbol{v}-\boldsymbol{K}_{Q_0},
\end{equation}
where $\boldsymbol{K}_{Q_0}$ is the vector of Riemann constants
\begin{equation}
\label{RC} (\boldsymbol{K}_{Q_0})_j=\dfrac{1+\Pi_{jj}}{2}
-\sum_{i=1,i\neq j}^g\oint_{\alpha_i}\d v_i(P)\int_{Q_0}^P\d v_j.
\end{equation}
Furthermore, the divisor $\sum_{i=1}^gQ_i$ is non special. 
\end{theorem}

For  a point $P\in \mathcal {C}$, we define the Abel map
$\boldsymbol{\mathfrak{A}}: \mathcal{C}\longrightarrow
\mathrm{Jac}(\mathcal {C})$ by setting
\begin{equation}
\label{AbelQ0}
\boldsymbol{\mathfrak{A}}(P)=\int_{Q_0}^P\d\boldsymbol{v},
\end{equation}
for some base point $Q_0\in\mathcal{C}$.
For a positive divisor $\mathcal{D}$ of degree $n$ the Abel map reads
\[
\boldsymbol{\mathfrak{A}}(\mathcal{D})=\int_{nQ_0}^{\mathcal{D}}\d\boldsymbol{v}.\]
There exists a non-positive  divisor $\Delta$ of degree $g-1$ such that
\begin{equation}
\label{VRC}
\boldsymbol{\mathfrak{A}}(\Delta-(g-1)Q_0)=\boldsymbol{K}_{Q_0},
\end{equation}
where $\boldsymbol{K}_{Q_0}$ has been defined in (\ref{RC}).
The divisor $\Delta$ is called the Riemann divisor and satisfies the condition
$2\Delta={\mathcal K}_{\mathcal C}$, where ${\mathcal K}_{\mathcal C}$ is the canonical class (that is the class of divisors of Abelian differentials).
The vector $\boldsymbol{e}$  defined in (\ref{vectore}) can be written in the form
\begin{equation}
\label{VRC0}
\boldsymbol{e}=\boldsymbol{\mathfrak{A}}(\sum_{i=1}^g Q_i-Q_0-\Delta).
\end{equation}

\section{Characteristics supported on branch points\label{charac}}
\noindent
In this section we are going to compute the characteristics $[\Uc_k]$
of the Abelian images of branch points $P_k=(\lb_k,0)$
\[
\Uc_k=\int_{P_{\infty}}^{P_k}\d\boldsymbol{v},\quad k=1,\ldots, 2m+1
,\quad P_{\infty}=(\infty,\infty),\]
 in terms of the period matrix $\Pi$.
\begin{lemma}\label{character}
The relations
\begin{align*}
&\int_{P_{2k}}^{P_{2k-1}}\d v_{k+sm}=\dfrac{N-1-s}{N},\quad
\int_{P_{2k+2}}^{P_{2k+1}}\d v_{k+sm}=-\dfrac{N-1-s}{N},\\
&\int_{P_{2k+2}}^{P_{2k+1}}\d v_{j+sm}=0,\;\;j\neq k,k+1,\;j=1,\dots,m,\\
&\int_{P_{2j}+1}^{P_{2j}}\d v_{k+sm}=\dfrac{N-1}{N}\Pi_{k+sm,j}-\dfrac{1}{N}
\sum_{r=1}^{N-2}\Pi_{k+sm,j+rm}
\end{align*}
are valid for $\quad k,j=1,\dots,m$, $s=0,\dots,N-2$.
\end{lemma}
The proof represents a generalization to the case $N>2$, of the
derivation of half-integer characteristics given e.g. in
\cite{fk80}; for detail see the proof of this lemma in \cite{eg04}.

From the relations given in the Lemma \ref{character} we are able
to write the characteristics $[\Uc_k]$ in the form
\begin{eqnarray*}
\label{U2m1}
&[\Uc_{2m+1}]&=\underbrace{\left[
\begin{array}{cccc}0&\ldots&0&\stackrel{m\downarrow}{0}\\
                   0&\ldots&0&\frac{1}{N}
\end{array}\right.}_m\dots
\underbrace{
\begin{array}{cccc}0&\ldots&0&\stackrel{sm\downarrow}{0}\\
                   0&\ldots&0&\frac{s}{N}\end{array}}_{m}\dots
\underbrace{\left.
\begin{array}{cccc}0&\ldots&0&\stackrel{(N-1)m\downarrow}{0}\\
                   0&\ldots&0&\frac{N-1}{N}\end{array}\right]}_{m},\\
\end{eqnarray*}
\begin{align*}
\label{U2m}
&[\Uc_{2m}]=\underbrace{\left[
\begin{array}{cccc}0&\ldots&0&\stackrel{m\downarrow}{-\frac{1}{N}}\\
                   0&\ldots&0&\frac{1}{N}
\end{array}\right.}_m\dots
\underbrace{
\begin{array}{cccc}0&\ldots&0&\stackrel{sm\downarrow}{-\frac{1}{N}}\\
                   0&\ldots&0&\frac{s}{N}\end{array}}_{m}\dots
\underbrace{\left.
\begin{array}{cccc}0&\ldots&0&\stackrel{(N-1)m\downarrow}{-\frac{1}{N}}\\
                   0&\ldots&0&\frac{N-1}{N}\end{array}\right]}_{m},\\
\nonumber
&\\
&~~~~~~~~~~~~~~~~~\qquad\qquad\qquad\vdots\\
&\\
&[\Uc_{2k+1}]=\underbrace{\left[
\begin{array}{cccccc}
0&\dots&\stackrel{k\downarrow}{0}&-\frac{1}{N}&\dots&-\frac{1}{N}\\
0&\ldots&\frac{1}{N}&0&\dots&0
\end{array}\right.}_m\dots
\underbrace{
\begin{array}{cccccc}
0&\dots&\stackrel{k+(s-1)m\downarrow}{0}&-\frac{1}{N}&\dots&-\frac{1}{N}\\
0&\ldots&\frac{s}{N}&0&\dots&0\end{array}}_{m}\dots\\
&~~~~~~~~~~~~~~~~~~~~~~~~~~~~~~~~~\qquad\qquad\quad\dots
\underbrace{\left.
\begin{array}{cccccc}
0&\dots&\stackrel{k+(N-2)m\downarrow}{0}&-\frac{1}{N}&\dots&-\frac{1}{N}\\
0&\ldots&\frac{N-1}{N}&0&\dots&0
\end{array}\right]}_{m},\\
&[\Uc_{2k}]=\underbrace{\left[
\begin{array}{cccccc}0&\dots&\stackrel{k\downarrow}{-\frac{1}{N}}&-\frac{1}{N}&\dots&-\frac{1}{N}\\
                   0&\ldots&\frac{1}{N}&0&\dots&0
\end{array}\right.}_m\dots
\underbrace{
\begin{array}{cccccc}0&\dots&\stackrel{k+(s-1)m\downarrow}{-\frac{1}{N}}&-\frac{1}{N}&\dots&-\frac{1}{N}\\
                   0&\ldots&\frac{s}{N}&0&\dots&0\end{array}}_{m}\dots\\
&~~~~~~~~~~~~~~~~~~~~~~~~~~~~~~~~~~~~~~~~~~~~~~~~\qquad\qquad\quad\dots
\underbrace{\left.
\begin{array}{cccccc}0&\dots&\stackrel{k+(N-2)m\downarrow}{-\frac{1}{N}}&-\frac{1}{N}&\dots&-\frac{1}{N}\\
                   0&\ldots&\frac{N-1}{N}&0&\dots&0\end{array}\right]}_{m},\\
&~~~~~~~~~~~~~~~~~~~~~~~~~~~~~~~~~~~~~~~~~~~~~~~~~~~\qquad\qquad\qquad\vdots\\
&[\Uc_{2}]=\underbrace{\left[
\begin{array}{cccc}-\frac{1}{N}&-\frac{1}{N}&\ldots&-\frac{1}{N}\\
                   \frac{1}{N}&0&\ldots&0
\end{array}\right.}_m\dots
\underbrace{
\begin{array}{cccc}\stackrel{1+(s-1)m\downarrow}{-\frac{1}{N}}&-\frac{1}{N}&\ldots&-\frac{1}{N}\\
                   \frac{s}{N}&0&\ldots&0\end{array}}_{m}\dots
\underbrace{\left.
\begin{array}{cccc}-\frac{1}{N}&-\frac{1}{N}&\ldots&-\frac{1}{N}\\
                   \frac{N-1}{N}&0&\ldots&0\end{array}\right]}_{m},\\
&[\Uc_{1}]=\underbrace{\left[
\begin{array}{cccc}-\frac{1}{N}&-\frac{1}{N}&\ldots&-\frac{1}{N}\\
                   0&0&\ldots&0
\end{array}\right.}_m\dots
\underbrace{
\begin{array}{cccc}-\frac{1}{N}&-\frac{1}{N}&\ldots&-\frac{1}{N}\\
                   0&0&\ldots&0\end{array}}_{m}\dots
\underbrace{\left.
\begin{array}{cccc}-\frac{1}{N}&-\frac{1}{N}&\ldots&-\frac{1}{N}\\
                   0&0&\ldots&0\end{array}\right]}_{m}.
\end{align*}

In the following we determine the vector of Riemann constants and 
Riemann divisor of the curve $\mathcal{C}$.
\begin{lemma}
The vector of Riemann constants computed in the homology basis described
in Figure~1 and  with  base point $P_{\infty}$
equals
\begin{equation}
\boldsymbol{K}_{\infty}=(N-1)\sum_{k=1}^m\int\limits_{P_{\infty}}^{P_{2k}}\d
\boldsymbol{v},\label{Rconstant}\quad P_{2k}=(\lb_{2k},0).
\end{equation}
The Riemann divisor $\Delta$  of the curve $\mathcal{C}$ in the homology basis described in Figure~\ref{fig2} is equivalent to
\begin{equation}
\Delta= (N-1)\sum_{k=1}^mP_{2k}- P_{\infty}.\label{riemanndiv}
 \end{equation}
\end{lemma}
\begin{proof}
The proof  of (\ref{Rconstant}) is obtained by direct calculations
from the definition (\ref{RC}) and Lemma~\ref{character}.
The relation (\ref{riemanndiv}) follows immediately from (\ref{Rconstant}).
\end{proof}

Following Diez \cite{diez91}, we  describe a family
of non-special divisors of degree $g$ on $\mathcal{C}$,
supported on the branch points.
Denote by $\boldsymbol{s}=(s_1,\ldots,s_{2m+1})$ a $2m+1$-vector with
non-negative entries $s_k$, satisfying the condition
\begin{align}
\sum_{i=1}^{2m+1} s_i=(N-1)m,\quad 0\leq s_i\leq N-1.
\end{align}
To each vector $\boldsymbol{s}$ we put into correspondence the  divisors
\begin{equation}  \mathcal{D}_{\boldsymbol{s}}=\sum_{k=1}^{2m+1}s_kP_k
\label{Dl} \end{equation}
where  $P_k=(\lb_k,0)$, $k=1\ldots,2m+1$, are  branch points.
In particular we shall consider the divisor class
$\mathcal{ D}_{\boldsymbol{m}}$ containing
$\left(\begin{array}{c}2m+1\\m\end{array}\right)$  divisors
\begin{equation}
\mathcal{D}_{\boldsymbol{m}}
=(N-1)P_{i_1}+\ldots+(N-1)P_{i_{m-1}}
+(N-1)P_{i_m},
\label{diez1}
\end{equation}
where the indices $\{i_1,i_2,\ldots,i_m\}\subset \{1,\ldots,2m+2\}$.
Among the divisors with $m+1$ branch points we consider the divisor class
$\mathcal{ D}_{\boldsymbol{m+1}}$ which contains
$\frac12(m+2)(m+1)\left(\begin{array}{c}2m+1\\m-1\end{array}\right)$ divisors
\begin{equation}
\mathcal{D}_{{\boldsymbol{m+1}}}=(N-1)P_{i_1}+
\ldots+(N-1)P_{i_{m-1}}+(N-2)P_{i_m}+P_{i_{m+1}},
\label{diez2}
\end{equation}
where the indices  $\{i_1,i_2,\ldots,i_m,i_{m+1}\}\subset \{1,\ldots,2m+2\}$.
It is out of the scope of the present manuscript to classify all the  
non-special divisors of the form (\ref{Dl}). 
However we can single out two families of non-special divisors.

\begin{lemma} \label{diezlemma}
The divisors $\mathcal{D}_{\boldsymbol{m}}$
defined in (\ref{diez1}) are non-special
 and the divisors
 $\mathcal{D}_{{\boldsymbol{m+1}}}$ defined in (\ref{diez2}) are non-special for $N>3$.
At $N=3$ the divisors
\begin{equation}
\label{diez22}
\mathcal{D}_{{\boldsymbol{m+1}}}=2P_{i_1}+
\ldots+2P_{i_{m-1}}+P_{i_m}+P_{i_{m+1}},
\end{equation}
are non-special if  one of the following conditions are satisfied
\begin{itemize}
\item $i_m$ and $i_{m+1}$ have different parity,
\item $i_m$ and $i_k$ have different parity for $k<m$, $m>1$,
\item $i_k$ and $i_j$ have different parity for $k,j<m$ and $m>1$.
\end{itemize} 
\end{lemma}
\begin{proof}
Assume the opposite: suppose that the divisor
$\mathcal{D}_{\boldsymbol{m}}$ or $\mathcal{D}_{\boldsymbol{m+1}}$ is special, 
this means that  there exists  a non-constant
meromorphic function $f(\lambda,\mu)$
whose divisor of poles is $\mathcal{D}_{\boldsymbol{m}}$ or $\mathcal{D}_{\boldsymbol{m+1}}$.
For simplify the proof, we assume that   $P_{i_k}\neq P_{\infty},\,k=1,\dots,m+1$.
Let $\mathbb{I}_l$ be a subset of $\{1,3,5,\dots,2m+1\}$ with $l$ distinct elements.
The function
$$\phi(\lambda,y)=f(\lambda,y)\prod_{i_j
\in \mathbb{I}_l}(\lambda-\lambda_{i_j}),\quad l=m,m+1,$$
has poles only at infinity. It follows from the {\it Weierstrass gap
theorem}, that the ring of meromorphic functions with
poles at infinity is generated in the case of the curve
$y^N=p(\lambda)q^{N-1}(\lambda)$ by powers of $\lambda$
and functions $y^k/q(\lambda)^{k-1}$, $k=1,\ldots,N-1$ with $q(\lambda)=
\prod_{j=1}^m(\lb-\lb_{2j})$.
Therefore the function  $\phi(\lambda,y)$  can be written  in the form
\begin{equation}\phi(\lambda,y)=R_0(\lb)+\sum_{k=1}^{N-1}R_k(\lambda)
\dfrac{y^k}{q^{k-1}(\lambda)},\label{crucial}
 \end{equation}
where $R_k(\lambda)$ are polynomials in $\lambda$.\footnote{In this point our proof differs from
  that  given in \cite{diez91} which is working for Galois covers of
  the form $y^N=\prod_{k=1}^{mN}(\lambda-\lambda_k) $ where the ansatz for
  the function (\ref{crucial}) can be written as $\sum R_iy^i$.  }

We remark that  $\mathrm{ord}_{\infty}\left(R_k(\lambda)
\dfrac{y^k}{q^{k-1}}\right)\neq
\mathrm{ord}_{\infty}\left(R_j(\lambda)\dfrac{y^j}{q^{j-1}}\right)$
for $k\neq j$ because otherwise
\begin{equation} N \mathrm{ord}_{\lambda} R_k(\lambda)+k = N\mathrm{ord}_{\lambda}R_j(\lambda)+j,
 \end{equation}
which implies $k=j$.
From this observation it follows  that
\begin{equation}
\mathrm{ord}_{\infty}(f(\lambda,y)\prod_{i_n\in \mathbb{I}_l}
(\lambda-\lambda_{i_n})  ) =
\mathrm{ord}_{\infty}\left(R_j(\lambda)\dfrac{y^j}{q^{j-1}(\lambda)} \right )
 \label{degree} \end{equation}
for some $0\leq j \leq N-1$.   Moreover
\[
\mathrm{ord}_{\infty}(f(\lambda,y)\prod_{i_n\in  \mathbb{I}_l}(\lambda-\lambda_{i_n})  )
=-N | \mathbb{I}_l|+k_l,\quad l=m,m+1,
  \]
where $k_l$, $l=m,m+1$, is the order at infinity of $f(\lambda,y)$ and
 $ | \mathbb{I}_l|=l$.
From the equation of the curve we get  $\mathrm{deg}\,
y=mN+1$. Therefore the equality (\ref{degree}) can be
written as
\[
N | \mathbb{I}_l|-k_l=N(r_j+m)+j,\quad l=m,\,m+1,
\]
where $r_j$ is the degree of $R_j(\lambda)$, so that
\[
j=N (| \mathbb{I}_l|-r_j-m)-k_l\geq 0
.\]
When $l=m$ that is  $ | \mathbb{I}_l|=m$,
it follow that $r_j=0, \,j=0,\, k_m=0$ and
\[
f(\lb,y)=\dfrac{1}{\prod_{i_n\in  \mathbb{I}_m}(\lambda-\lambda_{i_n})}
\] and contradicts the assumption that $f(\lb,y)$ has divisor $\mathcal{D}_m$.

When $l=m+1$, that is  $ | \mathbb{I}_l|=m+1$, two possibilities occurs:
(i) $r_j=0,\;j=N-k_{m+1},\;0\leq k_{m+1}<N$
and
(ii) $r_j=1,\,k_l=0,\;j=0$. This latter  case can be easily excluded 
while for the former one we have
\[
f(\lb,y)=\dfrac{1}{\prod_{i_j\in   \mathbb{I}_{m+1} }(\lambda-\lambda_{i_j})}
\dfrac{ y^{N-k_{m+1} }}{q^{N-k_{m+1}-1}(\lambda)} ,\quad
\mathrm{ord}_{\infty}(f(\lambda,y))=k_{m+1}
,\]
which has divisor
\[
\text{Div}f(\lb,y)=-N\sum_{i_n\in   \mathbb{I}_{m+1}}P_{i_n}+
(N-k_{m+1})\sum_{j=1}^{m+1}P_{2j+1}+k_{m+1}\sum_{j=1}^mP_{2j}.
\]
Namely the divisors of poles of $f(\lb,y)$ is
\[
\text{Div}_{\text{poles}}f(\lb,y)=(N-k_{m+1})\sum_{i_n\in
 \mathbb{I}_{m+1},i_n\text{ even}}P_{i_n}+k_{m+1}
\sum_{i_n\in   \mathbb{I}_{m+1},i_n\,\text{odd}}P_{i_n}
\]
and for $N>3$, differs from $\mathcal{D}_{\boldsymbol{m+1}}$.
This contradicts  the assumption unless $f$ is constant.
For $N=3$ the divisor of poles of $f(\lb,y)$ coincides with
$\mathcal{D}_{\boldsymbol{m+1}}$ in the following two cases:
\[
\mathcal{D}_{\boldsymbol{m+1}}=2\sum_{k=1}^{m-1}P_{i_k}+P_{i_m}+P_{i_{m+1}},
\]
with
\[
\;\;i_m,\,i_{m+1}\in\{2,4,6\dots,2m\},
\;\;\;i_k\in\{1,3,5,\dots,2m+1\},\;k=1,\dots,m-1
\]
or
\[i_m,\,i_{m+1}\in\{1,3,5\dots,2m+1\},
\;\;\;i_k\in\{2,4,6,\dots,2m\},\;k=1,\dots,m-1.
\]
We conclude that the divisors (\ref{diez22}) are non special when 
one of the following conditions are
satisfied: 1)
 $i_m$ and $i_{m+1}$ have
different parity;  2)  $i_m$ and $i_{k}$, $k<m$, $m>1$ 
 have different parity; 3)  $i_j$ and $i_{k}$ have
different parity for $j,k<m,\,m>1$.
\end{proof}

By the above lemma and by the Riemann singularity theorem~
\ref{vanishing},    the vector
\[
\boldsymbol{e_{\boldsymbol{m}}}=\boldsymbol{\mathfrak{A}}
(\mathcal{D}_{\boldsymbol{m}}-Q_0-\Delta),
\quad Q_0\notin  \mathcal{D}_{\boldsymbol{m}}.
\]
is a non singular $1/N$ period, namely 
$\theta(\boldsymbol{e_{\boldsymbol{m}}};\Pi)\neq 0.$
In the above expression we can get rid of the base point $Q_0$.
For the purpose, we introduce the divisor
\[
\mathcal{D}=Q+J(Q)+\dots+J^{N-1}Q,
\] 
which is independent from the point $Q$ and satisfies the 
relation
\[
\mathcal{D}\equiv NP_j,\quad j=1,\dots,2m+2, \;\;P_{2m+2}=P_{\infty}.
\]
Let $J_0=\{2,4,\dots,2m+2\}$ and $I_0=\{1,3,\dots,2m+1\}$ be a partition of 
the branch points in odd and even and  $P_{\infty}=P_{2m+2}$. 
Let $J_1\subset J_0$ and $I_1\subset I_0$ be a partition such that
\[
|J_1|+|I_1|=m+1.
\]
Consider the vector
\begin{equation}
\label{evector}
\boldsymbol{e_{\boldsymbol{m}}}=\boldsymbol{\mathfrak{A}}((N-1)\sum_{i\in I_1}P_i+(N-1)\sum_{j\in J_1}P_j-\mathcal{D}-\Delta).
\end{equation}
Then from Lemma~\ref{diezlemma},
the vector $\boldsymbol{e_{\boldsymbol{m}}}$ is non singular, namely $\theta(\boldsymbol{e_{\boldsymbol{m}}};\Pi)\neq 0.$
In the same way we define the vector obtained from the  divisors
$\mathcal{D}_{\boldsymbol{m+1}}$ in the following way
\begin{equation}
\label{evectorm}
\boldsymbol{e_{\boldsymbol{m+1}}}=\boldsymbol{\mathfrak{A}}((N-1)\sum_{i\in I_1}P_i+(N-1)\sum_{j\in J_1}P_j+(N-2)P_{i_m}+P_{i_{m+1}}-\mathcal{D}-\Delta),
\end{equation}
where now 
\[
|I_1|+|J_1|=m-1, \quad I_1\subset I_0,\;\; J_1\subset J_0,
\]
and 
\[
i_m\in (I_0-I_1),\quad i_{m+1}\in (J_0-J_1).
\]
From Lemma~\ref{diezlemma},
the vector $\boldsymbol{e_{\boldsymbol{m+1}}}$ is non singular.

In the next sections, we are going to study the Szeg\"o kernel associated to 
the characteristics (\ref{evector}) and (\ref{evectorm}).

\section{Kernel-forms}
The Schottky-Klein prime form $E(P,Q)$, $P,Q\in \mathcal{C}$ is a
skew-symmetric $(-\frac12,-\frac12)$-form on $\mathcal C\times\mathcal C$
\cite{fa73}
\begin{equation}
\label{prime} E(P,Q)=\frac{\theta[\boldsymbol{\gamma}]\left(\int\limits_P^{Q}\d
\boldsymbol{v};\Pi\right)}{ \sqrt{ \sum_{j=1}^g\frac{\partial }{\partial z_j}
\theta[\boldsymbol{\gamma}](\boldsymbol{0};\Pi) \d v_j(P)   }
 \sqrt{ \sum_{j=1}^g\frac{\partial }{\partial z_j}
\theta[\boldsymbol{\gamma}](\boldsymbol{0};\Pi) \d v_j(Q)   }},
\end{equation}
where $[\boldsymbol{\gamma}]$ are non-singular 
odd half-integer  characteristics.

The prime form does not depend on the characteristics $[\boldsymbol{\gamma}]$.
The automorphic factors of the prime form along all the cycles
$\alpha_k$ are trivial; the automorphic factor along each $\beta_k$
cycle in the $Q$ variable equals $ \exp\{-\pi \i \Pi_{kk}-2\pi
\i\int_{P}^Q\d v_k\}$. If the points $P$ and $Q$ are placed in the
vicinity of the point $R$ with local coordinate $x$, $x(R)=0$,
then the prime form has the following local behavior as
$Q\rightarrow P$
\begin{equation}
E(P,Q)=\frac{x(Q)-x(P)}{\sqrt{\mathrm{d}x(P)}\sqrt{\mathrm{d}x(Q)}
}
 \left(1+ O(1)) \right).
\end{equation}

The canonical bimeromorphic differential $\omega(P,Q)$ is defined as a 
symmetric 2-differential,
\begin{equation}
\omega(P,Q)=\mathrm{d}_{x(P)}\mathrm{d}_{x(Q)}\,\mathrm{log}\,
E(P,Q).\label{Bergmann}
\end{equation}
All $\alpha$-periods of $\omega(P,Q)$ with respect to any
of its  two variables vanish. The  period of the 2-differential $\omega(P,Q)$
with respect to the variable $P$  or $Q$,
along the $\beta_k$ cycle,
 is equal to $2\pi\i \d v_k(Q)$ or
$2\pi \i\d v_k(P)$ respectively.

The  2-differential $\omega(P,Q)$ has a double pole along the diagonal with the
following local behavior \cite{wi43,fa73}
\begin{align}
\omega(P,Q)&=\left(\frac{1}{(x(P)-x(Q))^2}+H(x(P),x(Q))
+\text{higher order terms}\right)\mathrm{d}x(P)\mathrm{d}x(Q),
\end{align}
where $H(x(P),x(Q))\mathrm{d}x(P)\mathrm{d}x(Q)$ is the non-singular part
of $\omega(P,Q)$ in each coordinate chart. The restriction
of $H$ on the diagonal is the Bergman projective connection 
(see for example \cite{tju78})
\begin{equation}
\label{pcon}
R(x(P))=6H(x(P),x(P)),
\end{equation}
which  depends non-trivially  on the chosen  system of local coordinate $x(P)$.
Namely the projective connection transforms as follows  with respect to a
change of local coordinates $x\rightarrow f(x)$
\[
R(x)\rightarrow R(f(x))[f'(x)]^2+\{f(x),x\},
\]
where $\{\,,\,\}$ is the Schwarzian derivative.

The Szeg\"o kernel $S\pq(P,Q)$ is defined for all non-singular
characteristics $\pq$ as the $(\frac12,\frac12)$-form on
$\mathcal{C}\times \mathcal{C}$ \cite{fa73}
\begin{equation}
\label{Szego} S\pq(P,Q)=\frac{\theta\pq\left(\int\limits^Q_{P} \d
\boldsymbol{v};\Pi \right)}{\theta\pq(\boldsymbol{0};\Pi)E(P,Q) }.
\end{equation}
The Szeg\"o kernel transforms when the variable
$Q$ goes around $\alpha_k$ and $\beta_k$-cycles as follows
\begin{align}
\label{Szego1}\begin{split}
&S\pq(P,\,Q+\alpha_k)=\e^{2\pi \i \delta_k}S\pq(P,Q), \\
&S\pq(P,\,Q+\beta_k)=\e^{-2\pi \i \epsilon_k}
S\pq(P,Q),\end{split}\quad  k=1,\ldots,g
\end{align}
The local behaviour of the Szeg\"o kernel when
$Q\rightarrow P$  is
\begin{equation}
\label{szegoexp}
S\pq(P,Q)=\dfrac{\sqrt{\mathrm{d}x(P)}\sqrt{\mathrm{d}x(Q)}}{x(Q)-x(P)}
\left[1+T(x(Q)) (x(Q)-x(P)) + O ((x(Q)-x(P)^2)\right],
\end{equation}
where 
\begin{equation}
\label{fayszego}
T(x(Q))dx(Q)=\sum_{k=1}^g\frac{\partial}{\partial z_k}
\log\theta\pq(\boldsymbol{0};\Pi)\mathrm{d}v_k(x(Q)).
\end{equation}
The important relation \cite{fa73}, Cor. 2.12, connects
the Szeg\"o kernel with canonical bimeromorphic differential
\begin{equation}
S[\boldsymbol{e}](P,Q)S[-\boldsymbol{e}](P,Q)=\omega(P,Q)
+\sum_{k,l=1}^g\frac{\partial^2}{\partial z_k\partial z_l}
\log\theta\pq(\boldsymbol{0};\Pi)\mathrm{d}v_k(P)\mathrm{d}v_l(Q).
\label{fay212}
\end{equation}
In the following we are going to give an algebraic expression for the
Szeg\"o kernel associated with  the characteristics
\[
\boldsymbol{e_{\boldsymbol{m}}}=\boldsymbol{\mathfrak{A}}((N-1)\sum_{i\in I_1}P_i+(N-1)\sum_{j\in J_1}P_j-\mathcal{D}-\Delta),
\]
defined in (\ref{evector}). For simplicity, we assume $2m+2\notin J_1$.
 We define the functions
\begin{equation}
\label{psi}
\psi_k(P,Q)=\frac{x(Q)-\lambda_k}{x(P)-\lambda_k},\quad k=1,\ldots,2m+1.
\end{equation}

\begin{theorem}
\label{tszegoN1}
The Szeg\"o kernel with characteristics 
$[\boldsymbol{e_{\boldsymbol{m}}}]$ defined  in (\ref{evector})
 is given by the formula
\begin{equation}
\label{szegoN1}
\begin{split}
S[\boldsymbol{e_{\boldsymbol{m}}}](P,Q)&=\dfrac{\sqrt{dx(P)dx(Q)}}{N(x(Q)-x(P))}\\&\times
\sum_{s=0}^{N-1}
\left[ \left( \dfrac{\prod_{i\in I_1} \psi_i}{\prod_{j\in J_2}\psi_j}\right)^{\frac{N-1}{2N}-\frac{s}{N} }
\left(\dfrac{\prod_{i\in I_2} \psi_i}{\prod_{j\in  J_1}\psi_j}\right)^{\frac{N-1}{2N}-
\left\{\frac{s+1}{N}\right\}}\right],
\end{split}
\end{equation}
where $\{\;\;\}$ is the fractional part and
\[
J_2=J_0-J_1-2m+2,\quad I_2=I_0-I_1.
\]
In particular, the Szeg\"o kernel with zero characteristics  is obtained
by fixing $I_1=I_0$ and  $J_1=\emptyset$ and takes the form
\begin{align}
\label{szegoN0} S[0](P,Q)&
=\dfrac{1}{N}\dfrac{ \sqrt{\mathrm{d}x(P)\mathrm{d}x(Q)}}{x(Q)-x(P)}
\sum\limits_{s=0}^{N-1} \left( \dfrac{q(x(P))}{p(x(P))}
\dfrac{p(x(Q))}{q(x(Q))} \right)^{-\frac{s}{N}+\frac{N-1}{2N}},
\end{align}
where the polynomials $p(\lambda)$ and $q(\lambda)$
have been defined in (\ref{pq}).
\end{theorem}
The proof is based on the uniqueness of the Szeg\"o kernel (see e.g.
Narasimhan \cite{nar87}) and the results of Nakayashiki  \cite{na97}. 
Indeed it is sufficient to check that the expression (\ref{szegoN1})  
is regular 
everywhere but on the diagonal where $P=Q$ and that its   divisor 
 in the variables $P$ and $Q$  coincides with the divisor 
of the Szeg\"o kernel given by the formula   (\ref{Szego}). 
Using the coordinate chart given in (\ref{localN}), the regularity of the 
expression 
(\ref{szegoN1}) can be checked in a straighforward way.
In the same way, by first fixing $P=P_{i_1}$ and then $Q=P_{i_1}$ with 
$i_1\in I_0\cup J_0 $ it is straighforward to obtain  the divisor class of 
(\ref{szegoN1}).
\begin{corollary}
\label{szegoexplemma}
The expansion of the  Szeg\"o kernel (\ref{szegoN1}) along
 the diagonal takes the form
\begin{equation}
\begin{split}
&S[\boldsymbol{e_{\boldsymbol{m}}}](P,Q)\simeq\frac{\sqrt{dx(P)dx(Q)}}{x(Q)-x(P)}\notag \\ 
& \times\left\{ 1+ \left(\frac{1}{12}\{x(Q),Q \} + \phi[\boldsymbol{e_{\boldsymbol{m}}}](Q)
\right)(x(P)-x(Q))^2\right\},
 \end{split}\label{szegoexpan1}
\end{equation}
where the function $\phi[\boldsymbol{e_{\boldsymbol{m}}}](\lb)$ takes the form
\begin{equation}
\begin{split}
\label{phi}
\phi[\boldsymbol{e_{\boldsymbol{m}}}](\lb)&=\frac{N^2-1}{24N^2}\left[\left(\dfrac{d}{d\lb}
\log\dfrac{\prod_{i\in I_1}(\lb-\lambda_i) }
{\prod_{j\in J_2}(\lb-\lambda_j)}\right)^2+
\left(\dfrac{d}{d\lb}\log\dfrac{\prod_{i\in I_2} (\lb-\lambda_i)}{\prod_{j\in  J_1}(\lb-\lambda_j)}\right)^2\right]\\
&+\frac{2(N-1)(N-5)}{24N^2}\left(\dfrac{d}{d\lb}\log\dfrac{\prod_{i\in I_1}(\lb-\lambda_i) }
{\prod_{j\in J_2}(\lb-\lambda_j)}\right)\left(\dfrac{d}{d\lb}\log\dfrac{\prod_{i\in I_2} (\lb-\lambda_i)}{\prod_{j\in  J_1}(\lb-\lambda_j)}\right).
\end{split}
\end{equation}
\end{corollary}
The proof of the above corollary is obtained by direct calculation.
We remark that for $I_1=I_0$ and $J_1=\emptyset$ the second term in (\ref{phi}) disappears
and we obtain
\[
\phi[\boldsymbol{0}](\lb)=\frac{N^2-1}{24N^2}\left(\dfrac{d}{d\lb}
\log\dfrac{p(\lb) }
{q(\lb)}\right)^2,
\]
where $p(\lb)$ and $q(\lb)$ have been defined in (\ref{pq}).

\noindent
Since 
\[
-\boldsymbol{e_{\boldsymbol{m}}}=\boldsymbol{\mathfrak{A}}((N-1)\sum_{i\in I_0-I_1}P_i+(N-1)\sum_{j\in J_0- J_1}P_j-\mathcal{D}-\Delta)
 \]
because
\[
(N-1)\sum_{k\in I_0\cup J_0}P_k-2\mathcal{D}-2\Delta\equiv 0
\]
we conclude that 
\[
\phi[\boldsymbol{e_{\boldsymbol{m}}}](Q)=\phi[-\boldsymbol{e_{\boldsymbol{m}}}](Q).
\]
Therefore, the following identity holds
\begin{equation}
\label{2szego}
\begin{split}
&S[\boldsymbol{e_{\boldsymbol{m}}}](P,Q)S[-\boldsymbol{e_{\boldsymbol{m}}}](P,Q)=\frac{\mathrm{d}x(P)\mathrm{d}x(Q)
}{(x(P)-x(Q))^2}\notag \\
&\times\left\{ 
1+ \left[\frac16 \{x(P),P \}+ 2\phi[\boldsymbol{e_{\boldsymbol{m}}}](P)\right](x(P)-x(Q))^2 +\ldots \right\}.
\end{split}
\end{equation}
The relations  (\ref{szegoexp}), (\ref{fayszego}) and the corollary~\ref{szegoexplemma}
imply
\begin{equation} 
\frac{\partial}{\partial z_k}
\theta[\boldsymbol{e_{\boldsymbol{m}}}](\boldsymbol{z};\Pi)\left|_{\boldsymbol{z}=0}=0\right.,\quad
k=1,\ldots,g,\label{thetader}
\end{equation}
for the vector $\boldsymbol{e_{\boldsymbol{m}}}$ defined in (\ref{evector}).

In the following we are going to give an algebraic expression for the
Szeg\"o kernel associated to   the characteristics
\begin{equation}
\label{evectorm1}
\boldsymbol{e_{\boldsymbol{m+1}}}=\boldsymbol{\mathfrak{A}}((N-1)\sum_{i\in I_1}P_i+(N-1)\sum_{j\in J_1}P_j+(N-2)P_{j_m}+P_{i_{m}}-\mathcal{D}-\Delta),
\end{equation}
where
\[
|I_1|+|J_1|=m-1,\quad i_m\in I_0-I_1,\;\;j_m\in J_0-J_1
\]
The construction is very similar to the previous case.
Also in this case, for simplifying the notation, 
we assume that $2m+2\notin J_1$ and $j_m\neq 2m+2$. We  define the following  sets
\[
I_2=I_0-I_1-i_{m},\;\;
J_2=J_0-J_1-j_{m}-2m+2.\;\
\]
\begin{theorem}
\label{tszegoN1m}
The Szeg\"o kernel with characteristics $[\boldsymbol{e_{\boldsymbol{m+1}}}]$ 
defined  in (\ref{evectorm1}) is given by the formula
\begin{equation}
\label{szegoN1m}
\begin{split}
S[\boldsymbol{e_{\boldsymbol{m+1}}}](P,Q)&
=\dfrac{\sqrt{dx(P)dx(Q)}}{N(x(Q)-x(P))}\times\\
&\sum_{s=0}^{N-1}
\left[ \left( \dfrac{\prod_{i\in I_1} 
\psi_i}{\prod_{j\in J_2}\psi_j}\right)^{\frac{N-1}{2N}-\frac{s}{N} }
\left(\dfrac{\prod_{i\in I_2} \psi_i}{\prod_{j\in  J_1}\psi_j}
\right)^{\frac{N-1}{2N}-
\left\{\frac{s+1}{N}\right\}}
\left(\dfrac{\psi_{i_m}}{\psi_{j_{m}}}\right)^{\frac{N-1}{2N}-
\left\{\frac{s+2}{N}\right\}}
\right],
\end{split}
\end{equation}
where $\{\;\;\}$ is the fractional part and the function $\psi_i$ has been defined in (\ref{psi}).
\end{theorem}
The proof is based on the uniqueness of the Szeg\"o kernel (see e.g.
Narasimhan \cite{nar87}) and the results of Nakayashiki  \cite{na97}. 
\begin{corollary}
\label{szegoexplemmam}
The expansion of the  Szeg\"o kernel (\ref{szegoN1}) along
 the diagonal takes the form
\begin{equation}
\begin{split}
&S[\boldsymbol{e_{\boldsymbol{m+1}}}](P,Q)\simeq\frac{\sqrt{dx(P)dx(Q)}}{x(Q)-x(P)}\notag \\ 
& \times\left\{ 1+ \left(\frac{1}{12}\{x(Q),Q \} + \phi[\boldsymbol{e_{\boldsymbol{m+1}}}](Q)
\right)(x(P)-x(Q))^2\right\},
 \label{szegoexpan1m}
\end{split}
\end{equation}
where the function $\phi[\boldsymbol{e_{\boldsymbol{m+1}}}](\lb)$ is 
\begin{equation}
\begin{split}
\label{phim}
\phi[\boldsymbol{e_{\boldsymbol{m+1}}}](\lb)&=\frac{N^2-1}{24N^2}\left[\left(\dfrac{d}{d\lb}
\log\dfrac{\prod_{i\in I_1}(\lb-\lambda_i) }
{\prod_{j\in J_2}(\lb-\lambda_j)}\right)^2+
\left(\dfrac{d}{d\lb}\log\dfrac{\prod_{i\in I_2} (\lb-\lambda_i)}{\prod_{j\in  J_1}(\lb-\lambda_j)}\right)^2+\right.\\
&+\left.
\left(\dfrac{d}{d\lb}
\log\dfrac{(\lb-\lambda_{i_m}) }
{(\lb-\lambda_{j_{m}})}\right)^2\right]\\
&+\frac{2(N-1)(N-5)}{24N^2}\left(\dfrac{d}{d\lb}\log\dfrac{\prod_{i\in I_1'}(\lb-\lambda_i) }
{\prod_{j\in J_2'}(\lb-\lambda_j)}\right)\left(\dfrac{d}{d\lb}\log\dfrac{\prod_{i\in I_2} (\lb-\lambda_i)}{\prod_{j\in  J_1}(\lb-\lambda_j)}\right)\\
&+\frac{2(N^2-12N+23)}{24N^2}\left(\dfrac{d}{d\lb}\log\dfrac{\prod_{i\in I_1}(\lb-\lambda_i) }
{\prod_{j\in J_2}(\lb-\lambda_j)}\right)\left(\dfrac{d}{d\lb}\log\dfrac{(\lb-\lambda_{i_m})}{(\lb-\lambda_{j_m})}\right),
\end{split}
\end{equation}
where
\[
I_1'=I_1+i_m,\;\;J_2'=J_2+j_m.
\]
\end{corollary}
The proof of the above corollary is obtained by direct calculation.

\noindent
Since 
\[
-\boldsymbol{e_{\boldsymbol{m+1}}}=\boldsymbol{\mathfrak{A}}(N-1)\sum_{i\in I_2}P_i+(N-1)\sum_{j\in J_0-J_1-j_m}P_j+(N-2)P_{i_m}-P_{j_m}-\mathcal{D}-\Delta)
 \]
because
\[
(N-1)\sum_{k\in I_0\cup J_0}P_k-2\mathcal{D}-2\Delta\equiv 0
\]
we conclude that 
\[
\phi[\boldsymbol{e_{\boldsymbol{m+1}}}](Q)=\phi[-\boldsymbol{e_{\boldsymbol{m+1}}}](Q).
\]
Therefore, the following identity holds
\begin{equation}
\label{2szegom}
\begin{split}
&S[\boldsymbol{e_{\boldsymbol{m+1}}}](P,Q)S[-\boldsymbol{e_{\boldsymbol{m+1}}}](P,Q)=\frac{\mathrm{d}x(P)\mathrm{d}x(Q)
}{(x(P)-x(Q))^2}\notag \\
&\times\left\{ 
1+ \left[\frac16 \{x(P),P \}+ 2\phi[\boldsymbol{e_{\boldsymbol{m+1}}}](P)\right](x(P)-x(Q))^2 +\ldots \right\},
\end{split}
\end{equation}
We remark that from (\ref{szegoexp}), (\ref{fayszego}) and the corollary~\ref{szegoexplemma}
\begin{equation} 
\frac{\partial}{\partial z_k}
\theta[\boldsymbol{e_{\boldsymbol{m+1}}}](\boldsymbol{z};\Pi)\left|_{\boldsymbol{z}=0}=0\right.,\quad
k=1,\ldots,g,\label{thetademr}
\end{equation}
for the characteristics $\boldsymbol{e_{\boldsymbol{m+1}}}$ defined in (\ref{evectorm}).
\section{Algebraic realization of the canoncal bimeromorphic differential}
The  canoncal bimeromorphic differential can be given in an algebraic form 
due to Klein
\cite{kl86},\cite{kl88} also \cite{ba97},\cite{hs66} and \cite{fa73}.
To develop this approach we first write the third kind differential
with poles in two arbitrary points. For the purpose we consider 
an arbitrary curve $\mathcal{C}$
 given by the polynomial equation $f(\lb,\mu)=0$ of degree $N$ in the variable $\mu$. We suppose that the curve $\mathcal{C}$ has a branch point at infinity. Let
\begin{equation} \boldsymbol{\Psi}(\lb,\mu)
=(1,\psi_1(\lb,\mu),\ldots,\psi_{N-1}(\lb,\mu))
\label{psibasis} \end{equation}
be the basis in the ring   $\mathcal{O}(\mathcal{C})$ of meromorphic 
functions on $\mathcal{C}$  with the only pole at infinity.
 There exists a  vector function
\begin{equation}
\boldsymbol{\Phi}(\lb,\mu)=(1,\Phi_1(\lb,\mu),\ldots,\Phi_{N-1}(\lb,\mu))
\label{phibasis} \end{equation}
for which
\begin{equation} \langle\boldsymbol{\Psi}(\lb,\mu),
\boldsymbol{\Phi}(\lb,\mu')  \rangle
=\frac{f(\lb,\mu')-f(\lb,\mu)}{\mu'-\mu}.
\end{equation}

Let $Q=(\lambda^{\prime},\mu^{\prime})$ and $R
=(\lambda^{\prime\prime},\mu^{\prime\prime})$ be two arbitrary
points of the curve $\mathcal C$. The third kind differential
$\Omega_{Q,R}(P)$ with
simple poles with residues $\pm 1$ in the points $P=Q$
and $P=R$ is given by the formula
\begin{equation}\label{3kdiff}
\Omega_{Q,R}(P)
=\left(\frac {\langle \boldsymbol{\Psi}(Q),
\boldsymbol{\Phi}(P)   \rangle   }{\lambda-\lambda^{\prime} }
  - \frac{   \langle \boldsymbol{\Psi}(R),
\boldsymbol{\Phi}(P)\rangle}
{\lambda-\lambda^{\prime\prime} }\right)\frac{\d \lambda}
{f_{\mu}(\lb,\mu) },
\end{equation}
with $f_{\mu}(\lambda,\mu) = \partial f(\lambda,\mu)/\partial \mu $.
In the case of the curve (\ref{CNM}) the vectors
$\boldsymbol{\Psi}(\lambda,\mu)$  (see the proof of Lemma \ref{diezlemma} )
and $\boldsymbol{\Phi}(\lambda,\mu)$ take the form
\begin{align}
\boldsymbol{\Psi}(\lambda,\mu)&=\left(1,\mu,\frac{\mu^2}{q(\lambda)},\ldots,
\frac{\mu^{N-1}}{q(\lambda)^{N-2}}\right),\label{psivector}\\
\boldsymbol{\Phi}(\lambda,\mu)&=\left( \mu^{N-1},\mu^{N-2},
q(\lambda)\mu^{N-3},\ldots, q(\lambda)^{N-2}  \right).
\label{phivector}
\end{align}

The canoncal bimeromorphic differential can be obtained by differentiation 
of (\ref{3kdiff})
as follows
\begin{equation}\label{kleinbergmann}
\omega(P,Q)=
\d \lambda'\frac {\partial}{\partial \lambda^{\prime}}
 \frac{\langle \boldsymbol{\Psi}(Q),
\boldsymbol{\Phi} (P) \rangle}{\lambda-\lambda'}\frac{\d \lambda}
{f_{\mu}(\lb,\mu)  }  +  \d \chi(P,Q),
\end{equation}
where $\d\chi(P,Q)$ is the  2-form  uniquely defined by the requirement that
(\ref{kleinbergmann}) is a symmetric  bi-differential normalized with respect to the $\alpha$ cycles and with the  only pole of second order along the diagonal.
The above arguments lead to the following algebraic expression for the
canoncal bimeromorphic differential:
\begin{align}
\label{algbergmann}
\omega(P,Q)&=
\frac{\partial}{\partial \lambda'}\frac{1}{\lambda-\lambda'}
\left[1+\sum_{s=1}^{N-1}\frac{{\mu'}^s 
q(\lambda)^{s-1}}{\mu^sq(\lambda')^{s-1}}
 \right]\frac{\mathrm{d}\lambda\mathrm{d}\lb'}{N}
+\d\chi(P,Q),\\
\label{chi}
\d\chi(P,Q)&=-\frac{1}{N}\sum_{s=1}^N\sum_{j=1}^m\lambda^{j-1}
\frac{q(\lb)^{s-1}}{\mu^s} \frac{q(\lb')^{N-s-1}}{{\mu'}^{N-s}}
\mathcal{R}_{s,j}(\lb')
\mathrm{d}\lb \mathrm{d}\lb',
\end{align}
with certain polynomials $\mathcal{R}_{s,j}(\lb,\mu)$. Here we do not need the exact form
of these polynomials.

Expanding the above expression along the diagonal, we obtain 
the Bergman projective connection 
\begin{align}
{\mathcal H}(P)&=\frac16 \{ x(P),P \}+
\frac{\d\chi(P,P)}{(\d x(P))^2}\notag\\
  &-\frac14 \frac{N-1}{N}\left( \frac{\frac{\partial^2}{\partial \lb^2} p(\lb)}
{p(\lb)} +
  \frac{\frac{\partial^2}{\partial \lb^2} q(\lb)}{q(\lb)} \right)
 \label{kleincon}
+\frac{N^2-1}{12N^2}\left( \frac{\d}{\d \lb}\,\mathrm{log}\,
\frac{p(\lb)}{q(\lb)}\right)^2.
\end{align}
Combining the above relation with the expansion (\ref{2szego}) of 
the Szeg\"o kernel 
along the  diagonal and the Fay relation (\ref{fay212}) which connects 
the canoncal bimeromorphic differential and Szeg\"o kernel,  we obtain 
an algebraic expression 
for the second 
derivatives of the theta function, namely
\begin{equation}
\label{second}
\begin{split}
&\dfrac{1}{(\d x(P))^2}\sum_{k,l=1}^g\frac{\partial^2}{\partial z_k\partial z_l}
\log\theta [\boldsymbol{e_{\boldsymbol{m}}}]  (\boldsymbol{0};\Pi)\mathrm{d}v_k(P)\mathrm{d}v_l(P)\\
&=\frac14 \frac{N-1}{N}\left( \frac{\frac{\partial^2}{\partial \lb^2} p(\lb)}
{p(\lb)} +
  \frac{\frac{\partial^2}{\partial \lb^2} q(\lb)}{q(\lb)} \right)-\frac{N^2-1}{12N^2}\left( \frac{\d}{\d \lb}\,\mathrm{log}\,
\frac{p(\lb)}{q(\lb)}\right)^2 +2\phi[\boldsymbol{e_{\boldsymbol{m}}}](P),
\end{split}
\end{equation}
where $\phi[\boldsymbol{e_{\boldsymbol{m}}}](P)$ has been defined in (\ref{phi}).

The following relation will be useful for the proof of the Thomae type
formula. It connects the function $\d\chi(P,Q)$ defined in (\ref{chi}) 
with the derivative with respect to the branch points of the determinant 
of the matrix $\mathcal{A}$ of $\alpha$-periods. 
\begin{lemma}\label{aslemma} For $s=1,\ldots,N-1$ the
following identities are valid
\begin{equation}
\frac{\partial }{\partial \lb_i}\;\mathrm{log}\;\mathrm{det}\;\mathcal{A}_s=
\frac{1}{\prod\limits_{l=1,l\neq i}^{2m+1}(\lambda_i-\lambda_l)}
\sum_{j=1}^m\lambda_{i}^{j-1}\mathcal{R}_{s,j}
(\lambda_i),\quad i=1,\ldots,2m+1.\label{logas}
\end{equation}
\end{lemma}
\begin{proof}
We integrate $\omega(P,Q)$ in the variable $P$ along the $\alpha_k$-cycle and
expand $\oint_{\alpha_k}\omega(P,Q)=0$ in the variable
$Q$ in the vicinity of the branch point $(\lambda_i,0)$ where the local
coordinate is introduced as $x(Q)=\lambda_i+\xi^N$,
$i\in\{1,\ldots,2m+1\}$. In this way we obtain for every fixed $s$
the rule of variation with respect to the  branch points of the first line of
$\mathcal{A}_s$-block of the period matrix $\mathcal{A}$:
\[ \frac{\partial}{\partial \lb_i}(\mathcal{A}_s)_{1,k}=\frac{1}
{\prod\limits_{l=1,l\neq i}^{2m+1}(\lambda_i-\lambda_l)}\sum_{j=1}^m\mathcal{R}_{s,j}
 (\mathcal{A}_s)_{j,k},
\quad k=1,\ldots,m.    \]
The derivative of all other lines can be obtained from the equivalence
\[  \frac{\partial}{\partial \lb_i}(\mathcal{A}_s)_{n,k}=\lambda_i^{n-1}
 \frac{\partial}{\partial \lb_i}(\mathcal{A}_s)_{1,k}+\frac{1}{N}
\sum_{l=1}^{n-1}(\mathcal{A}_s)_{l,k}\lambda_i^{n-l-1},
\quad n=2,\ldots,m,\; k=1,\ldots,m . \]
Therefore  one can write
\[ \frac{\partial}{\partial \lb_i}(\mathcal{A}_s)=\mathcal{P}_s(\lambda_i)
\mathcal{A}_s,\quad \mathrm{Tr}\,\mathcal{P}
=\sum_{j=1}^m\lambda_i^{j-1}\mathcal{R}_{s,j}(\lambda_i).  \]
The equality (\ref{logas}) follows.
\end{proof}
\begin{proposition}
\label{propder}
The following relation is valid
\begin{equation}\label{logdet}
\frac{\partial}{\partial
\lb_i}\,\mathrm{log}\,\mathrm{det}\,\mathcal{A} = -\res[P=P_i]
\frac{\d\chi(P,P)}{(\d x(P))^2}.
\end{equation}
\end{proposition}
\begin{proof}
The residue of (\ref{chi}) is
\[
\frac{1}{\prod\limits_{l=1,l\neq
i}^{2m+1}(\lambda_i-\lambda_l)}\sum_{s=1}^{N-1}
\sum_{j=1}^m\lambda_{i}^{j-1}\mathcal{R}_{s,j}(\lambda_i) .\] Then  use
the decomposition (\ref{Aperiod}) and the Lemma \ref{aslemma}.
\end{proof}
\begin{example}In the case of elliptic functions the formula (\ref{logdet})
represents the known relation
\[ \frac{\partial \omega}{\partial \lambda_i}=-\frac12\frac{\eta+\lambda_i\omega}
{(\lambda_{i}-\lambda_j)(\lambda_i-\lambda_k)}, \quad i\neq j\neq k=1,2,3,  \]
where the Weierstrass notations are used.
\end{example}

\section{Derivation of the Thomae type formula}
Now we are in a  position to derive Thomae type formula
\begin{theorem}\label{thomaezver}
Let $[\boldsymbol{e_{\boldsymbol{m}}}]$ be nonsingular $1/N$ 
characteristics,  given by the formula
\[
\boldsymbol{e_{\boldsymbol{m}}}=\boldsymbol{\mathfrak{A}}((N-1)\sum_{i\in I_1}P_i+(N-1)\sum_{j\in J_1}P_j-\mathcal{D}-\Delta),
\]
where
 $J_1\subset J_0=\{2,4,\dots,2m+2\}$ and $I_1\subset I_0=\{1,3,\dots,2m+1\}$, with 
\[
|J_1|+|I_1|=m+1.
\]
The Thomae type formula takes the form
\begin{equation}
\begin{split}
\label{thomaefinal}
\theta[\boldsymbol{e_{\boldsymbol{m}}}](0;\Pi)^{4N}&
=\dfrac{\prod_{i=1}^{N-1} \det \mathcal{A}^{2N}_i}{(2\pi\i)^{2mN(N-1)}}
\prod_{1\leq i<k\leq m }(\lambda_{2i}-\lambda_{2k})^{N(N-1)}\prod_{0\leq i<k\leq m }
(\lambda_{2i+1}-\lambda_{2k+1})^{N(N-1)}\\
&\times\left(\dfrac{\prod_{i\in I_1,j\in J_1}(\lambda_i-\lambda_j)
\prod_{i\in I_2,j\in J_2}(\lambda_i-\lambda_j)}
{\prod_{i\in I_1,k\in I_2}(\lambda_i-\lambda_k)
\prod_{j\in J_1,k\in J_2}(\lambda_j-\lambda_k)}\right)^{2(N-1)},
\end{split}
\end{equation}
where
\[
I_2=I_0-I_1,\quad J_2=J_0-J_1-2m+2.
\]
When $I_1=I_0$ and $J_1=\emptyset$, the above formula reduces to 
\begin{equation}
\label{thomaefinal0}
\theta[\boldsymbol{0}](0;\Pi)^{4N}=\dfrac{\prod_{i=1}^{N-1} \det \mathcal{A}^{2N}_i}{(2\pi\i)^{2mN(N-1)}}
\prod_{1\leq i<k\leq m}(\lambda_{2i}-\lambda_{2k})^{N(N-1)}\prod_{0\leq i<k\leq m}(\lambda_{2i+1}-\lambda_{2k+1})^{N(N-1)}.
\end{equation}
\end{theorem}
\begin{proof}
Using the heat equation (\ref{heat}),
the Rauch variational formula (\ref{Rauch1}) and (\ref{thetader}) we have
\begin{align}
&\frac{\partial}{\partial \lb_i}\,\mathrm{log}\theta
 [\boldsymbol{e_{\boldsymbol{m}}}](\boldsymbol{0};\Pi)=\sum_{k,r=1}^{(N-1)m} \frac{\partial}{\partial \Pi_{k,r} }\,\mathrm{log} \,\theta
[\boldsymbol{e_{\boldsymbol{m}}}](\boldsymbol{0};\Pi)\frac{\partial
\Pi_{k,r}}{\partial
\lb_i}\label{thomae1}\\
&= \frac12\res[P=(\lb_i,0)]\left\{\dfrac{1}{(\d x(P))^2}
\sum_{k,r=1}^{(N-1)m} \frac{\partial^2}{\partial z_k\partial z_r}
\mathrm{log}\, \theta[\boldsymbol{e_{\boldsymbol{m}}}](\boldsymbol{0};\Pi)
\sum_{s=1}^N\d v_k(P^{(s)}) \d v_r(P^{(s)})\right\}.\notag
\end{align}

To proceed we shall represent the residue as a logarithmic
derivative. The equality  (\ref{second}) and the proposition~\ref{propder}
 enable us to  compute the residue in the r.h.s of (\ref{thomae1}) obtaining 
\begin{align*}
&\frac{\partial}{\partial \lb_i}\,\mathrm{log}\theta
 [\boldsymbol{e_{\boldsymbol{m}}}](\boldsymbol{0};\Pi)\\
&=\frac{N-1}{8} \res[\lb=\lb_i] \left( \frac{\frac{\partial^2}
  {\partial \lb^2} p(\lb)}{p(\lb)} +
  \frac{\frac{\partial^2}{\partial \lb^2} q(\lb)}{q(\lb)} \right)
  +\dfrac{1}{2}\frac{\partial}{\partial \lb_i} \,\mathrm{log}\,\mathrm{det}
\mathcal{A} \label{thomae2} \\
&+\res[P=(\lb_i,0)]\phi[\boldsymbol{e_{\boldsymbol{m}}}](P)-\frac{N^2-1}{24N}
\res[\lb=\lb_i]\left( \frac{\d}{\d
\lb}\,\mathrm{log}\,\frac{p(\lb)}{q(\lb)}
  \right)^2,
\end{align*}
where $\phi[\boldsymbol{e_{\boldsymbol{m}}}](P)$ is defined in (\ref{phi}).

Let us compute each residue in the above formula: 
\begin{align*} &\res[\lb=\lb_{2i}]  \frac{\frac{\partial^2}
  {\partial \lb^2} p(\lb)}{p(\lb)} =2 \frac{\partial}{\partial \lb_{2i}} \,
\mathrm{log}
  \prod\limits_{\begin{array}{c}1\leq i< k\leq m\end{array}}
  (\lb_{2i}-\lambda_{2k})\, \text{and}\; \res[\lb=\lb_{2i+1}] 
 \frac{\frac{\partial^2}
  {\partial \lb^2} p(\lb)}{p(\lb)}=0,\\
&\res[\lb=\lb_{2i+1}]  \frac{\frac{\partial^2}
  {\partial \lb^2} q(\lb)}{q(\lb)} =
  2 \frac{\partial}{\partial \lb_{2i+1}} \,\mathrm{log}
  \prod\limits_{\begin{array}{c}1\leq i<k\leq m\end{array}}
  (\lb_{2i+1}-\lambda_{2k+1})\; \text{and}\; \res[\lb=\lb_{2i}] 
 \frac{\frac{\partial^2}
  {\partial \lb^2} q(\lb)}{q(\lb)}=0.
  \end{align*}
Also we have for any partition $  {I}\cup   {J}$ and any $\lb_n$
\begin{align*}
\res[\lb=\lb_n]\left( \frac {\mathrm{ d}}{\mathrm{d}
\lb}\,\mathrm {log} \frac { \prod\limits_{i\in   {I}}
(\lb-\lambda_{i}) }{ \prod\limits_{j\in   {J}} (\lb-\lambda_{j}) }
\right)^2=2 \frac{\partial}{\partial \lb_{n}} \; \mathrm{log}
\frac{\prod\limits_{i<k\in {I}} (\lb_i-\lambda_{k})
\prod\limits_{i<k\in {J}} (\lb_i-\lambda_{k})
}{\prod\limits_{ i\in {I}, j\in {J} }
(\lb_i-\lb_{j})}.
\end{align*}
Furthermore
\begin{align*}
\res[\lb=\lb_n]&\left[\left(\dfrac{d}{d\lb}\log
\dfrac{\prod_{i\in I_1}(\lb-\lambda_i) }
{\prod_{j\in J_2}(\lb-\lambda_j)}\right)\left(\dfrac{d}{d\lb}
\log\dfrac{\prod_{i\in I_2} (\lb-\lambda_i)}{\prod_{j\in  J_1}
(\lb-\lambda_j)}\right)\right]\\
&=\dfrac{\partial}{\partial \lb_n}\log
\left(\dfrac{\prod_{i\in I_1, k \in I_2}(\lambda_i-\lambda_k)
\prod_{j\in J_1 k\in J_2}(\lambda_j-\lambda_k)}
{\prod_{i\in I_1,j\in J_1}(\lambda_i-\lambda_j)
\prod_{i\in I_2,j\in J_2}(\lambda_i-\lambda_j)}\right).
\end{align*}
The integration in $\lambda_n$ then gives
\begin{equation}
\begin{split}
\label{thomaeintegr}
\theta[\boldsymbol{e_{\boldsymbol{m}}}](0;\Pi)^{4N}&=C^{2N}\prod_{i=1}^{N-1} 
\det \mathcal{A}^{2N}_i\prod_{1\leq i<k\leq m}(\lambda_{2i}-
\lambda_{2k})^{N(N-1)}\prod_{0\leq i<k\leq m}(\lambda_{2i+1}
-\lambda_{2k+1})^{N(N-1)}\\
&\times\left(\dfrac{\prod_{i\in I_1,j\in J_1}(\lambda_i-\lambda_j)
\prod_{i\in I_2,j\in J_2}(\lambda_i-\lambda_j)}
{\prod_{i\in I_1,k\in I_2}(\lambda_i-\lambda_k)
\prod_{j\in J_1,k\in J_2}(\lambda_j-\lambda_k)}\right)^{2(N-1)},
\end{split}
\end{equation}
where $C$ is a constant independent of $\Pi$ and the  branch points.
To derive the above relation we also used the decomposition of the period 
matrix 
$\mathcal{A}$ into blocks $\mathcal{A}_i$. 

To compute $C$ we shall use the original Thomae arguments \cite{th69}.
We consider the Thomae type formula (\ref{thomaeintegr}) at zero 
characteristics, that is, the partition $I_1=I_0$ and $J_1=\emptyset$ and 
pinch the branch points in the following way
\[
\lb_{2k}=v_k-\epsilon,\quad \lb_{2k-1}=v_k +\epsilon \quad
k=1,\dots,m,\quad 0<\epsilon\ll 1.
\]
In this case the l.h.s of (\ref{thomaeintegr}) becomes
 $\theta[\boldsymbol{0}](\boldsymbol{0};\Pi)=1+O(\epsilon)$.
Regarding the r.h.s the following relations are needed:
\[
\lim_{\epsilon\rightarrow 0}(\mathcal{A}_s)_{i,j}=
2\pi\i\dfrac{v_i^{j-1}}{\prod\limits_{\substack{k\neq
i\\k=1}}^m(v_i-v_k) (v_i-\lambda_{2m+1})^{\frac{s}{N}}}
\]
so that
\begin{equation}
\label{lim0} \lim_{\epsilon\rightarrow 0}(\det {\mathcal
A}_s)=\left(2\pi \i\right)^m\dfrac{1}{\prod \limits_{\substack{k<
j\\k,j=1}}^m(v_k-v_j)} \dfrac{1}{\prod\limits_{k=1}^m
(v_k-v_{2m+1})^{\frac{s}{N}}}.
\end{equation} 
Combining the above relations the expression for the constant  $C=(2\pi \i)^{-m(N-1)}$.
\end{proof}

It is straightforward to check that for $N=2$ the formula
(\ref{thomaefinal}) coincides with the original Thomae formula
(\ref{thomaeoriginal}).

To complete the section we give the Thomae formula for  the characteristics
$[\boldsymbol{e_{\boldsymbol{m+1}}}]$ defined in (\ref{evectorm}).
\begin{theorem}
Let $[\boldsymbol{e_{\boldsymbol{m+1}}}]$ are nonsingular $1/N$ 
characteristics,  given by the formula
\[
\boldsymbol{e_{\boldsymbol{m+1}}}=\boldsymbol{\mathfrak{A}}((N-1)\sum_{i\in I_1}P_i+(N-1)\sum_{j\in J_1}P_j+(N-2)P_{j_m}+P_{i_m}-\mathcal{D}-\Delta),
\]
where
 $J_1\subset J_0=\{2,4,\dots,2m+2\}-\{2m+2\}$ and $I_1\subset I_0=\{1,3,\dots,2m+1\}$, with 
\[
|J_1|+|I_1|=m-1
\]
and $j_m\in J_0-J_1-2m+2$, $i_m\in I_0-I_1$.
The Thomae type formula takes the form
\begin{equation}
\begin{split}
\label{thomaefinalm}
\theta[\boldsymbol{e_{\boldsymbol{m+1}}}](0;\Pi)^{4N}&
=\dfrac{\prod_{i=1}^{N-1} \det \mathcal{A}^{2N}_i}{(2\pi)^{2mN(N-1)}}
\prod_{1\leq i<k\leq m}(\lambda_{2i}
-\lambda_{2k})^{N(N-1)}\prod_{0\leq i<k\leq m}(\lambda_{2i+1}
-\lambda_{2k+1})^{N(N-1)}\\
&\times\left(\dfrac{\prod_{i\in I_1',j\in J_1}(\lambda_i-\lambda_j)
\prod_{i\in I_2,j\in J_2'}(\lambda_i-\lambda_j)}
{\prod_{i\in I_1',k\in I_2}(\lambda_i-\lambda_k)
\prod_{j\in J_1,k\in J_2'}
(\lambda_j-\lambda_k)}\right)^{2(N-1)}\\
&\times\left(\dfrac{\prod_{j\in J_2}(\lambda_{i_m}-\lambda_j)
\prod_{i\in I_1}(\lambda_i-\lambda_{j_m})}
{\prod_{i\in I_1}(\lambda_{i_m}-\lambda_{i})
\prod_{j\in J_2}(\lambda_{j_m}-\lambda_j)}\right)^{4(N-2)},
\end{split}
\end{equation}
where 
\[
I_1'=I_1+i_m,\quad I_2=I_0-I_1-i_m,\;\;J_2=J_0-J_1-j_m-2m+2,\;\;J_2'=J_2+j_m.
\]
\end{theorem}
The proof of the above formula follows the steps of the proof 
of (\ref{thomaefinal}).

\begin{example} We consider as an example the trigonal curve of genus 
two (studied 
by Hutchinson \cite{hut02}, see also \cite{eg04} for more details)
\begin{equation}
\mu^3=(\lambda-\lambda_1)(\lambda-\lambda_3)(\lambda-\lambda_2)^2.
\label{hutcurve}
\end{equation}

In the homology basis given in Figure~\ref{fig2}, the  Riemann period matrix is
of the form 
\begin{equation}
\Pi=\left(\begin{array}{cc} 2T&T\\T&2T   \end{array} \right), \quad 
T=\frac{\imath\sqrt{3}}{3}\frac{F\left(\frac13,\frac23;1;1-t\right)}
{F\left(\frac13,\frac23;1;t\right)},
\end{equation}
where $F(a,b,c;t)$ is standard hypergeometric function and 
$t=(\lambda_2-\lambda_1)/(\lambda_3-\lambda_1)$.
In the homology basis given in figure~\ref{fig2}  the characteristics of the  branch points are
\begin{align*}
[\mathfrak{A}_1]=\left[ \begin{array}{cc} 
\frac23&\frac23\\0&0\end{array}  \right],\quad 
[\mathfrak{A}_2]=\left[ \begin{array}{cc} 
\frac23&\frac23\\\frac13&\frac23\end{array}  \right],\quad 
[\mathfrak{A}_3]=\left[ \begin{array}{cc} 
0&0\\\frac13&\frac23\end{array}  \right], \quad [\mathfrak{A}_4]=[0],
\end{align*}
whilst characteristics of the vector of Riemann constants are
$[ \boldsymbol{K}_{P_{\infty}}  ]=2[\mathfrak{A}_2] $.  
Given
\[
I_0=\{1,3\},\quad J_0=\{2,4\}, \quad P_4=P_{\infty},
\]
we introduce the following  partitions of the branch points.
\begin{enumerate}
\item The partition
\[
I_1=\{1,3\}, \;J_1=\emptyset \quad \text{or}\quad I_1=\emptyset,\;J_1=\{2,4\},
\]
 which corresponds to the characteristics  $[\boldsymbol{e_{\boldsymbol{m}}}]= [\boldsymbol{0}]$.
\item The partition
\[
I_1=\{1\},\;\;J_1=\{2\},
\]
or
\[
I_1=\emptyset,\quad J_1=\emptyset, \;\;i_m=3,\;j_m=2,
\]
which corresponds to the  characteristics  $ [\boldsymbol{e_{\boldsymbol{m}}}]=\left[ \begin{array}{cc} 
\frac13&\frac13\\0&0\end{array}  \right]$.
\item  The partition 
\[
I_1=\{3\},\;\;J_1={2},
\]
or
\[
I_1=\emptyset,\quad J_1=\emptyset,\;\; i_m=1,\;j_m=2.
\]
which corresponds to the characteristics $[ \boldsymbol{e_{\boldsymbol{m}}}]=\left[ \begin{array}{cc} 
0&0\\\frac23&\frac13\end{array}  \right]$.
\end{enumerate}
In the first case, the Thomae formula takes the form
\[ 
\theta[\boldsymbol{0}](\boldsymbol{0};\Pi)^4=\frac{\mathcal{A}_1^2
\mathcal{A}_2^2}{16\pi^4}(\lambda_1-\lambda_3)^2,   
\]
where
\begin{align}
\mathcal{A}_1=\frac{2\sqrt{3}\pi}{3}\frac{1-\rho^2}
{(\lambda_3-\lambda_1)^{1/3}}
F\left(\frac13,\frac23;1;t,\right), \quad \mathcal{A}_2=-\rho\mathcal{A}_1, 
\end{align}
where $\rho^3=1$. The second partition gives 
\[ \theta\left[\begin{array}{cc}
\frac13&\frac13\\0&0 \end{array}\right](\boldsymbol{0};\Pi)^{12}
=\left(\frac{\mathcal{A}_1
\mathcal{A}_2}{4\pi^2}\right)^6(\lambda_1-\lambda_2)^{4}
(\lambda_1-\lambda_3)^{2}  
\]
and the third partition
\[ \theta\left[\begin{array}{cc} 0&0\\
\frac23&\frac13 \end{array}\right](\boldsymbol{0};\Pi)^{12}
=\left(\frac{\mathcal{A}_1
\mathcal{A}_2}{4\pi^2}\right)^6(\lambda_3-\lambda_2)^{4}
(\lambda_3-\lambda_1)^{2}. 
\]
The quotient of the first and third relations leads to 
\[
\left(\frac{\theta\left[\begin{array}{cc} 0&0\\
\frac23&\frac13 \end{array}\right](\boldsymbol{0};\Pi)}
{\theta(\boldsymbol{0};\Pi)}\right)^6=\dfrac{(\lambda_3-\lambda_2)^2}{(\lambda_3-\lambda_1)^2},
\]
which we mentioned in \cite{eg04}.
\end{example}

\section{Conclusion}
In this paper we presented the derivation of the Thomae type formula 
for singular curves with $Z_N$ symmetry.
We computed explicitly the Riemann period matrix of the curve
in the fixed homology basis given in figure~\ref{fig2}
taking into account the action of automorphism.
We also described $1/N$ periods of the curve in terms of its characteristics
given in the form of $g\times2$-matrices with rational entries.

We considered in the papers a family of non-special divisors supported
on the  branch points and derived  the Thomae formula for this family only.
But our derivation is general and is working for any other family of
non-singular $1/N$-periods. The proof given above goes up to the original
Thomae proof and involved a number of steps such as Rauch variational
formulae, calculation of the holomorphic projective connection, 
certain results about the Bergman kernel.
But the key-point of the proof is the derivation of an
algebraic expression for Szeg\"o kernel associated to the  
aforementioned 1/N-period. 
We show that the values of the exponents in the Thomae formula follows 
directly from this expression.

The results of the paper can be generalized to the family of curves
\[ \mu^N=(\lambda-\lambda_1)^{m_1}\ldots (\lambda-\lambda_k)^{m_k}, \quad
m_1+\ldots+m_k \quad \text{divisible by}\quad N.  \]

However even for the curve considered  a number of open problems remain,
among which the  complete classification of non-special divisors 
 supported on the  branch points and the derivation of an algebraic
expressions for the associated Szeg\"o kernels. 
We remark that a family which is different  from the one  discussed in the present
manuscript  has already been studied in \cite{eg04}.

Another set of interesting problems consist of the derivation of
Thomae formulae for $\theta$-derivatives, which generalize the  formulae
given by Thomae in the case of hyperelliptic curve \cite{th69}.
Such formulae are important for obtaining the   expressions of the 
 ``winding vectors'' in terms of $\theta$-constants and 
 which are generalization of the Rosenhain
formulae known for genus two.

We can also point out the  problem of the derivation for the given curve
of the Jacobi-Riemann derivative formula, one of the most mysterious
$\theta$-formula, which Riemann called as a ``pearl of mathematics''.

The answers to these questions are of interest for the applied problems
pointed out in the Introduction. The investigation of the particular
curve undertaken above was stimulated by its links with the solvable
Riemann-Hilbert problem of rank $N$ considered in \cite{eg04}.


\end{document}